\documentstyle[epsfig,referee]{mn2e}
\def\H1743{H~1743-322}

\title[The last three outbursts of \H1743 by RXTE]
  {The last three outbursts of H1743-322 observed by RXTE in its latest service phase}
\author[J. N. Zhou et al.]
  {J.~N.~Zhou$^{1,2,3}$\thanks{E-mail: zjn@pmo.ac.cn},
   Q.~Z.~Liu$^{1,3}$, Y.~P.~Chen$^{2}$, J.~Li$^{2}$, J.~L.~Qu$^{2}$, S.~Zhang$^{2}$
  \newauthor
  H.~Q.~Gao$^{2}$ and Z.~Zhang$^{2}$\\
  $^{1}$ Purple Mountain Observatory, Chinese Academy of Sciences, Nanjing 210008, China\\
  $^{2}$ Laboratory for Particle Astrophysics, Institute of High Energy Physics, Beijing 100049, China\\
  $^{3}$ Key Laboratory of Dark Matter and Space Astronomy, Chinese Academy of Sciences, Nanjing 210008, China\\}

\begin{document}

\label{firstpage}

\maketitle

\begin{abstract}
After 2010, three outbursts of \H1743 were detected by {\it RXTE}. We have carried out timing and spectral analysis of the data, emphasizing on the two with relatively complete evolution history presented in the {\it RXTE}/PCA observations. We then constitute an enlarged outburst sample for \H1743 which allows to investigate the spectral transitions in more details. We find that the spectral transitions to high-soft state constrain a region for four outbursts in hardness-intensity diagram. An extension of the region locates in the vicinity of the failed outburst in 2008, and excludes it from a successful group. We therefore suggest the failed outburst in 2008 may present the first almost successful outburst sample, which is important for modelling the outburst, especially upon the conditions required for transition to high-soft state.
\end{abstract}

\begin{keywords}
 X-rays: binaries -- X-rays: stars -- X-rays: individual: IGR
J17464-3213 -- \H1743 -- black hole candidate
\end{keywords}

\section{Introduction}

Black Hole Transients (BHTs) are characterized by the transitions among spectral states: Low-Hard State (LHS), Hard-Intermediate State (HIMS), Soft-Intermediate State (SIMS) and High-Soft State (HSS) (see e.g. Belloni 2011). BHTs spend most time in quiescent state. During an outburst, the evolution of transitions forms a q-track in hardness-intensity diagram (HID), and the luminosity in X-ray band varies, in general, by several orders of magnitude. State transitions occur not only in X-ray band, but also in other wavelengths, e.g. radio, IR and optical (Homan et al. 2005; Lalemci et al. 2005; Fender et al. 1999).

Different states lie in different areas in HID. LHS corresponds to the right vertical branch in HID. The X-ray spectrum is hard and dominated by a power-law component (see e.g. Belloni et al. 2011, Motta et al. 2009). The power density spectrum (PDS) shows a strong variability ($\sim$ 30\% fractional rms) in the form of a band-limited noise. Intermediate states, including Hard-Intermediate State (HIMS) and Soft-Intermediate State (SIMS), occur after LHS. In this phase, the spectrum is softer and gradually dominated by a combination of thermal and non-thermal components. The PDS presents an obvious broad peak (sometimes with a Lorentzian shape) apart from the weaker band-limited noise. If an outburst reaches HSS, the spectrum is dominated by a soft thermal component, accompanying with a weak steep power-law component. In this phase, little variability would be observed.

A QPO ({\it Quasi-Periodic Oscillation}) is defined as the broad peak above the continuum when performing a power spectrum of the X-ray time series, and has been discovered in many systems. QPOs are thought to originate from the innermost regions of the accretion flows around the stellar-mass black holes (Motta et al. 2011). Low-frequency QPOs with frequency ranging from mHz to a few Hz are a common feature in almost all BH X-ray binaries. According to the quality factor (Q = $\upsilon_{centroid}$/FWHM) and the shape of the noise, QPOs can be classified into type A, B and C (Casella et al. 2004, Motta et al. 2011). Type-C QPO, which can be identified in the PDS with two main components: a strong flat-top noise and one or more QPO peaks, usually are observed in late LHS and HIMS. Comparatively, type-B QPO with a weak noise and is typical to SIMS. Type-A QPO usually shows broad peak (Q$\leq$3) associated with a weak red noise. There is no QPOs or sometimes with type-A QPOs in HSS(Belloni et al. 2011).

The black hole candidate X-ray binary \H1743 was discovered in August 1977 with HEAO-1 (Doxsey et al. 1977) and Ariel 5 (Kaluzienski \& Holt 1977). Ten outbursts since 2003 have been observed by {\it RXTE}.

In March 2003, \H1743 was detected in outburst with the {\it International Gamma-Ray Astrophysics Laboratory} ({\it INTEGRAL}) (Revnivtsev et al. 2003) and {\it RXTE} (Markwardt \& Swank 2003). This outburst is the brightest one ever since its first detection, up to $\sim$1.3 Crab at 1.5-12 keV. The radio (Rupen et al. 2003), infrared (Baba et al. 2003) and optical (Steeghs et al. 2003) observations were followed quickly. Steiner et al. (2011) applied a symmetric kinematic model to constrain the parameters, using {\it RXTE} data in 2003 together with radio bands. The distance (inclination angle) of this source was estimated to 8.5$\pm$0.8 kpc (75$^{\circ}\pm$3$^{\circ}$). The mass of the black hole is inferred to be 6.1 M$_{\odot}$ to 23 M$_{\odot}$ and its spin is 0.2$\pm$0.3 (68\% confidence level).

The following two fainter outbursts (September 2004 and September 2005) were not reported since few observations were available, especially for the former one. In January 2008, another outburst was detected by {\it RXTE}/ASM ({\it All-Sky Monitor}) (Kalemci et al. 2008). During this outburst, the source reached the HSS showing a canonical q-track. A secondary outburst in the same year was detected by {\it INTEGRAL} (Kuulkers et al. 2008) in September and the source was in the hard state. {\it Swift} and {\it RXTE} monitoring observations helped following the spectral evolution of the outburst. On October 23 (MJD 54762) the observation by {\it Swift} indicated that the source had undergone a state transition from LHS to HIMS (Belloni et al. 2008). During this outburst, the softest point had a hardness ($\sim$0.5), much larger than previous outbursts (Capitanio et al. 2009). It is the so-called failed outburst, a premature one in the absence of HSS due to the decrease of the mass accretion rate. Chen et al. (2010) found that an equal-division line in HID can be used  to distinguish this failed outburst from the others. The next outburst of \H1743 was detected on July 2009, by {\it Swift}/BAT ({\it Burst Alert Telescope}) (Krimm et al. 2009). {\it VLA} ({\it Very Large Array}) caught the simultaneous radio activity. In this outburst, Miller-Jones et al. (2012) reported the possible disc-jet coupling.

\begin{figure}
\centering
	\includegraphics[width=0.5\textwidth]{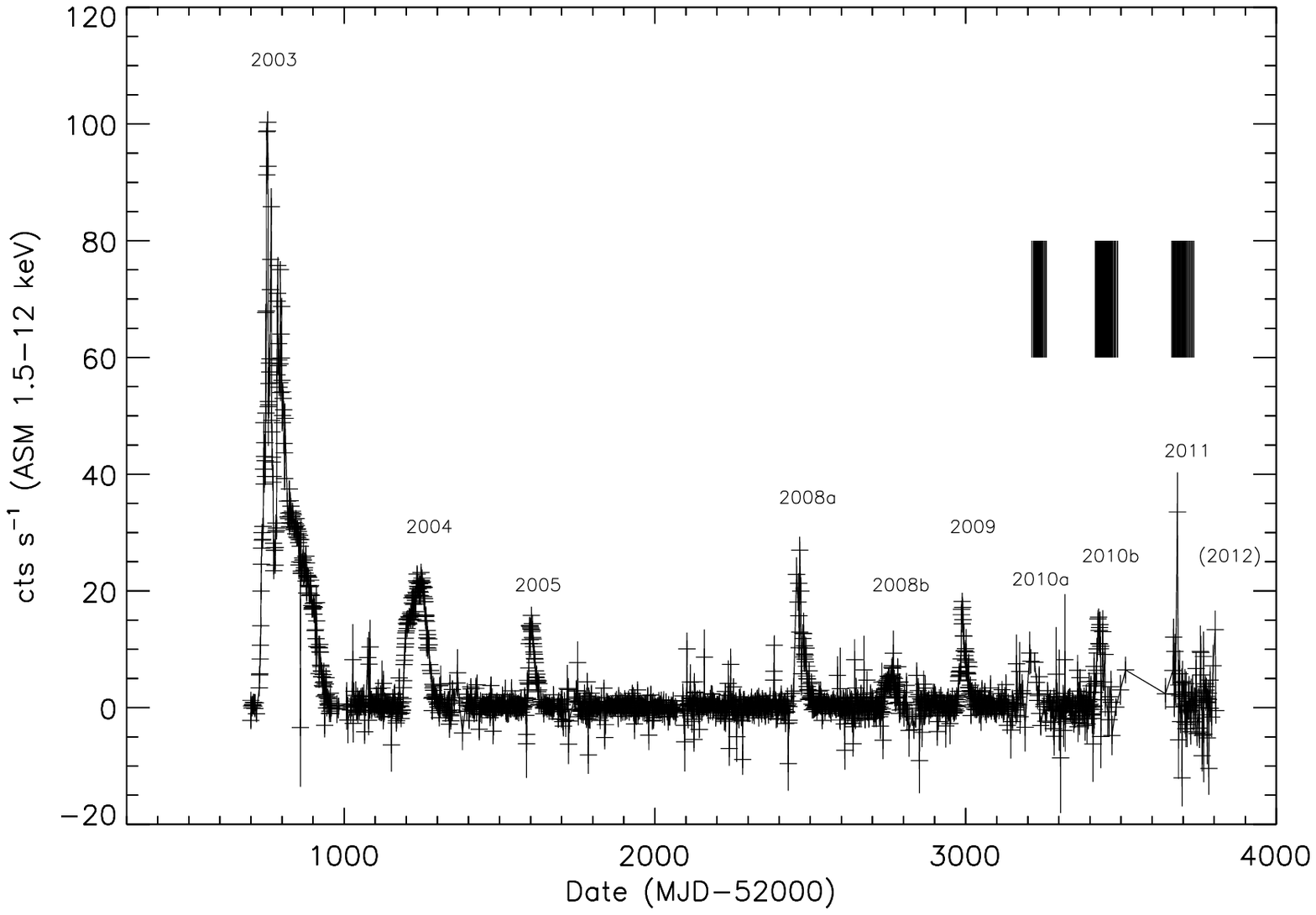}
		\caption{The {\it RXTE}/ASM light curve of ten outbursts of \H1743. The vertical lines (three block regions) indicate the times that we used for {\it RXTE} pointings.}
	\label{lcs}
\end{figure}

On 27 December, 2009 (MJD 55172), {\it MAXI}/GSC detected a new outburst of \H1743 (Yamaoka et al. 2009) and it was observed by {\it RXTE} since 17 January 2010. There were two more outbursts observed by {\it RXTE} afterwards on August 2010 (MJD 55417) and April 2011 (MJD 55663). We name these three outbursts in this paper as 2010a, 2010b and 2011, respectively. Although {\it Swift}/BAT (see Fig.~\ref{lcs}) detected another outburst in December 2011, {\it RXTE} was decommissioned formally on January 5, 2012 and no data are available any longer. Therefore, the 2010a, 2010b and 2011 outbursts are the last three outbursts recorded by {\it RXTE} in its   latest  service phase.

In this paper, we carry out a systematic timing and spectral analysis on the last three outbursts of \H1743 observed by {\it RXTE}. We report the outburst evolution and compare them to those recorded by {\it RXTE} previously.

\section{OBSERVATION AND ANALYSIS}

The {\it RXTE} observation for these three outbursts started on MJD 55213 for 2010a (33 pointed observations), MJD 55417 for 2010b (58 observations) and  MJD 55663 for 2011 (39 observations). Table 1-3 list all the observations included in this work. HEAsoft (version 6.11.1) is used for data analysis. Since combining large spectral data sets with different PCA ({\it Proportional Counter Array}) configurations (with associated differing calibrations) can produce large systematic errors, it is preferable to use a single common PCU ({\it Proportional Counter Unit}) for all available data sets (Smith et al. 2009). PCU2 (in the 0-4 numbering scheme) is used for most of the analysis as it is the only PCU that was 100$\% $ on during the observation. Deadtime correction is not needed because of the very low count rate ($<$ 500 cts/sec/PCU for almost all observations). We select the time intervals with the constraints of \textit{ELV} $>$ 10$^{\circ}$ and \textit{OFFSET} $<$ 0.02$^{\circ}$, where the \textit{ELV} is the angle above the Earth limb and the \textit{OFFSET} is the angular distance between the pointing and the source. PCA background light curves and spectra are generated with FTOOLS task {\it pcabackest}. The tool {\it pcarsp} is adopted to generate PCA response matrices for spectra. For the background file  we choose the most recent one available from the HEASARC's website\footnote{http:$//$heasarc.gsfc.nasa.gov$/$docs$/$xte$/$pca$\_$news.html} for the bright sources.

For spectral analysis, we take into account the 3-30 keV data of PCA. All the spectra and light curves from Standard-2 data are produced with FTOOLS tool {\it rex} script\footnote{http:$//$heasarc.gsfc.nasa.gov$/$docs$/$xte$/$recipes$/$rex.html}, which is designed to be a convenient tool for doing the same basic analysis on a large amount of data, available at HEASARC's website. Our spectral model consist of absorption (``WABS" in XSPEC), a multicolor disk blackbody (``DISKBB"), a power-law (``POWERLAW") and a gaussian (``GAUSS") for fitting iron line, which is the widely used in modelling black hole candidate (BHC) sources (Kalemci et al. 2006). The hydrogen column density is fixed to \textit{n}$_{\rm H}$=1.8$\times$10$^{22}$ cm$^{-2}$ (Coriat et al. 2011). An additional 0.5\% of systematic error is added in our spectral analysis with XSPEC v12.7.0. The model parameters are estimated at 90\% confidence level.

For the timing analysis, we compute the power density spectra (PDS) for each observation from the PCA event data with a time resolution of 125 $\mu$s in three energy bands, including 3-6 keV, 6-15 keV, and 15-30 keV. We also compute the PDS in the energy rage of 3-30 keV (Kalemci et al. 2006). For PDS, the power density is normalized to the squared mean intensity as described in Miyamoto et al. (1991). We fitted all PDSs with 2-3 Lorentzians for locating QPOs (frequency, FWHM, rms). The parameters were estimated at 90\% confidence level. The fractional rms was integrated over the frequency from 0.1 to 32 Hz.

\section{RESULTS}

\subsection{Timing Analysis}

\begin{figure}
\centering
	\includegraphics[width=0.5\textwidth]{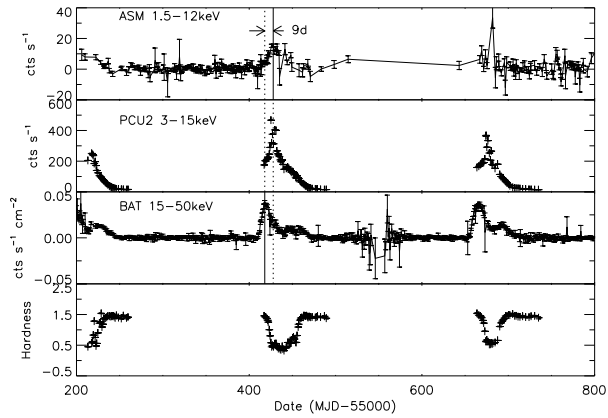}
		\caption{ASM, PCA \& BAT light curves of the three new outbursts adopted in this research. The hardness is defined as the ratio of the count rates between 6-15 keV band and 3-6 keV band of PCA.}
	\label{3lc}
\end{figure}

In Fig.~\ref{3lc} we show the light curves of the latest three outbursts of \H1743 in different energy ranges. The hard X-ray (BAT 15-50 keV) peak precedes the soft X-ray (ASM 1.5-12 keV) peak by about 9 days. Similar lag was also observed in previous outburst of \H1743 (see Chen et al. 2010) as well as in other sources (e.g. XTE J1550-564, Aql X-1, 4U 1705-44, see Yu et al. 2004). For Swift J1539.2-6227, the lag is about 8.5 days (Krimm et al. 2011). Yu et al. (2004) discussed this effect, suggesting a special structure for the accretion geometry.

\begin{figure}
\centering
\includegraphics[scale=0.5]{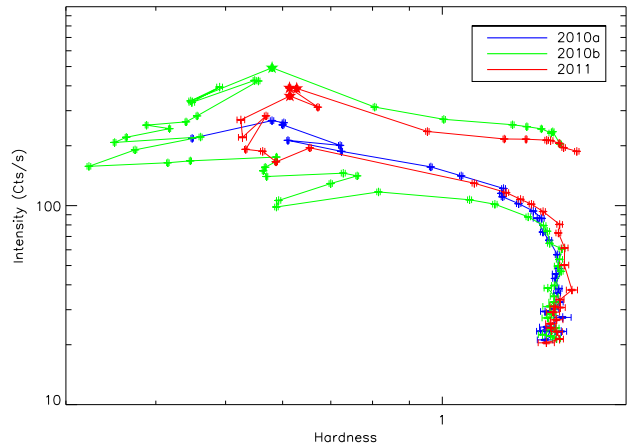}
\caption{The hardness-intensity diagram of the last three  outbursts observed by {\it RXTE}/PCA. The three energy bands, 3-6 keV, 6-15 keV and 15-30 keV, were used here. The hardness was defined with the ratio of the count rate of the first two bands, and the intensity is the total count rates of all three bands. The blue one is the 2010a outburst lacking the early observations. The five point star stand for the type-B QPOs.}
\label{3hids}
\end{figure}

\begin{figure*}
\centering
	\includegraphics[width=0.9\textwidth]{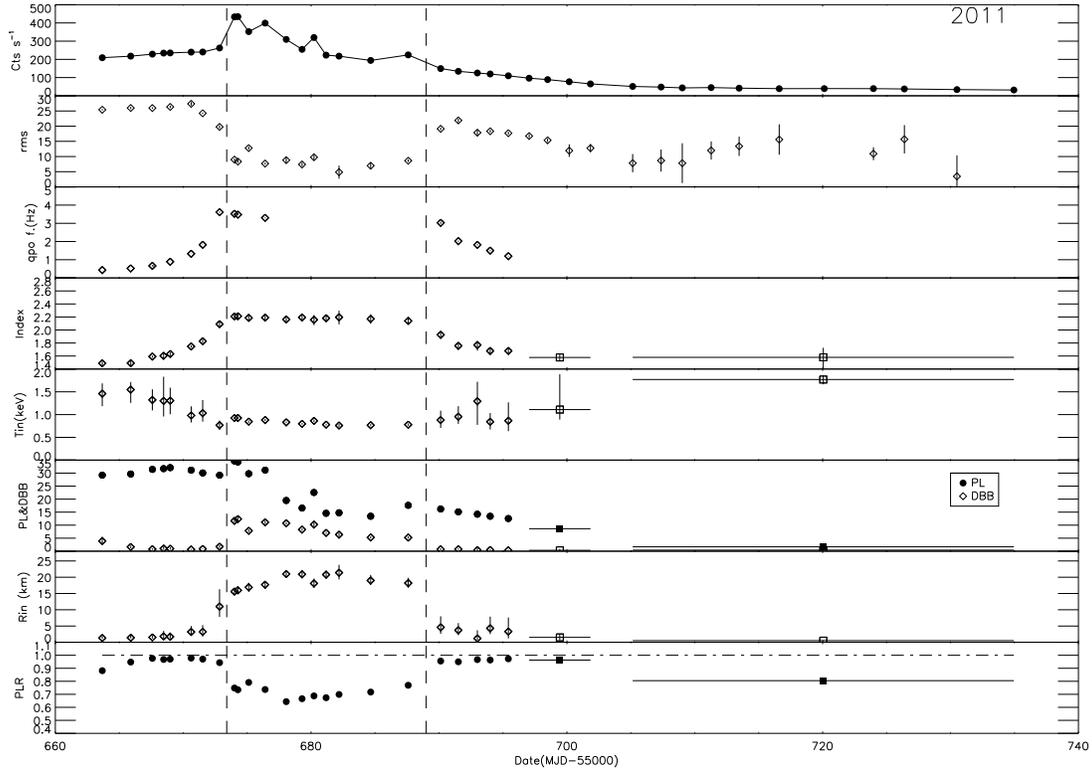}
		\caption{The fitted parameters of the last three outbursts 2010a, 2010b and 2011. In each panel, both timing and spectral evolution are presented. Eight sub-panel describe the light curve of PCU2, 0.1-32 HZ fractional rms amplitude, the QPO frequency, the photon index, the disk temperature, the disk blackbody flux and the power-law flux in 3-30 keV band in units of 10$^{-10}$ ergs cm$^{-2}$ s$^{-1}$, the reduced inner radius of the disk (R$_{in}$=D(10kpc)$\times$N$^{1/2}$), the ratio of the power-law flux to the total flux in 3-30 keV band, respectively. The vertical dash lines indicate the transitions between the hard-intermediate state and the thermal dominate state.}
	\label{mjdplot2}
\end{figure*}

The HIDs of three outbursts are compared in Fig.~\ref{3hids}. {\it RXTE} only covers the decay phase of the 2010a outburst but almost the full cycles for the other two outbursts (2010b and 2011). Although there was no observation in the early time of the 2010a outburst, the QPO absence and the low fractional rms (Fig.~\ref{mjdplot2}) implied that the first {\it RXTE} coverage was in HSS, and in the decay phase (Belloni et al. 2011). The first observation by {\it RXTE} of the second outburst was on MJD 55417. The source was in HIMS after initial rise (Fig.~\ref{3hids}). In the following evolution, the source moved horizontally to the softer area, the hardness peaking at almost $\sim$0.3, then slowly returned to the hard track. The third outburst tracked the same evolution, but presented an even larger hardness at the softest point. Therefore, two of these three outbursts provide complete evolution into our sample.

\begin{figure}
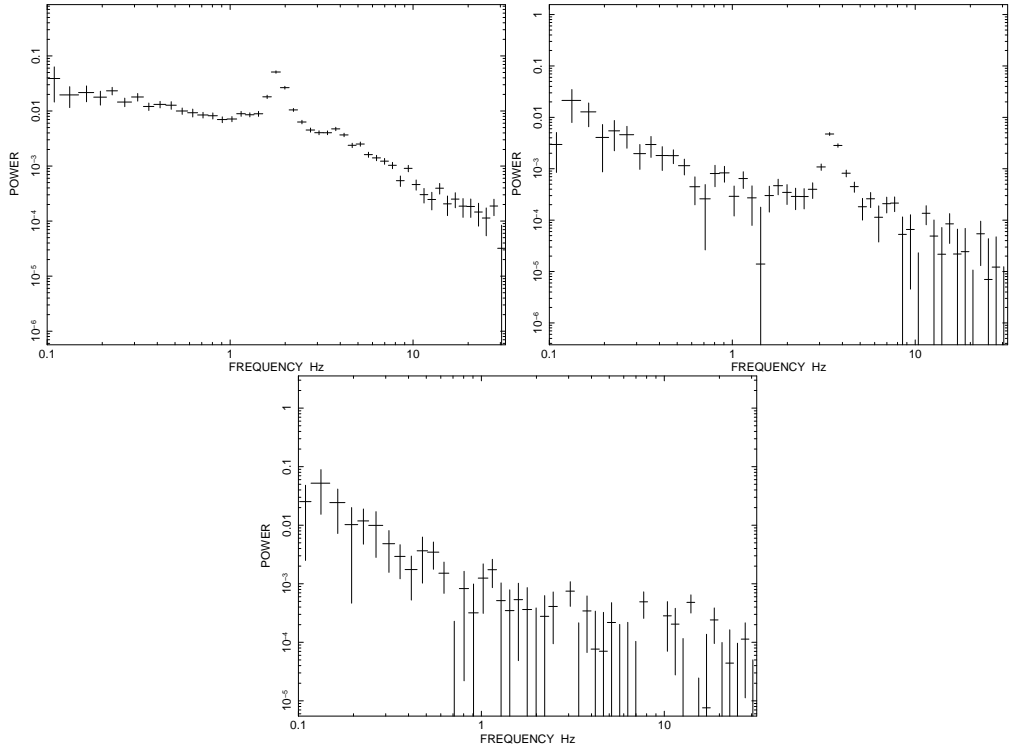

\centering
	\includegraphics[angle=-90 , width=0.4\textwidth]{HIMS.ps}
	\includegraphics[angle=-90 , width=0.4\textwidth]{SIMS.ps}
	\includegraphics[angle=-90 , width=0.4\textwidth]{HSS.ps}
		\caption{The power-density spectrum plot in different states, which derived from the 2011 outburst. (\textit{Top panel}: hard-intermediate state, \textit{Middle panel}: soft-intermediate state, \textit{Bottom panel}: high-soft state.)}
	\label{pdsplot}
\end{figure}

\begin{figure}
\centering
\includegraphics[scale=0.5]{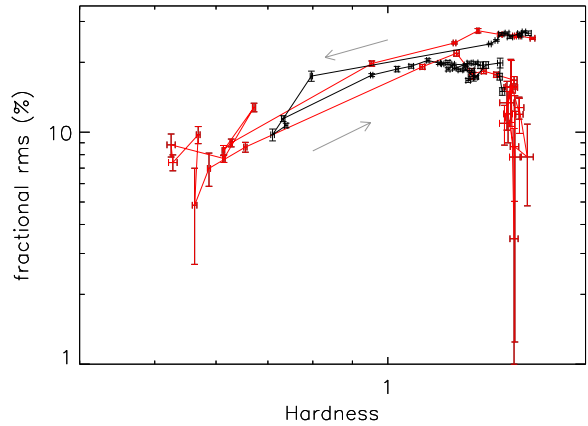}
\caption{The hardness-rms diagrams (HRDs) of 2008b (black) and 2011 (red) outbursts. They evolve along the direction of the arrows. During the 2011 outburst it is obviously these points are divided into three groups, with the left group standing  for the HSS.}
\label{hrd}
\end{figure}

Fig.~\ref{pdsplot} shows three typical power density spectra during the 2011 outburst. The upper left panel shows a significant flat-top noise (red noise) plus a QPO peak. According to the classification of QPOs and their corresponding spectral states, it is in HIMS. The upper right panel belongs to SIMS for the lack of red noise. The lower one has no QPO signal and belongs to HSS. We also show the hardness-rms relation of the 2011 outburst in Fig.~\ref{hrd}. Three distinguished regions are useful to discriminate  the transition states. The left region is at a low level of variability, for which the fractional rms (integrating the normalized   PDSs from 0.1 to 32 Hz) is below 10\%, suggesting that the 2011 outburst had reached to HSS. The right two regions denote the LHS/HIMS at the rise phase (right top) and decay phase (right down), respectively. The arrows show the evolution in time.

\subsection{Spectra Analysis}

In our fitting the column density is fixed to 1.8$\times$10$^{22}$ cm$^{-2}$ and an iron line at 6.4 keV is assumed. The reduced chi-squares of fits fluctuates between 0.527 and 1.427. All the parameters found in the fits are presented in Table 1-3 and Fig.~\ref{mjdplot2}. The main results for the three outbursts are summarized in the following.

(\textit{a}) In the first outburst, after the break (MJD 55222, vertical dash line), the fractional rms increased significantly, accompanying with an appearance of QPO feature and a hardened spectrum ($\Gamma$ $<$ 2.1). The thermal emission also started to decay to a low level, and non-thermal emission gradually dominated the luminosity. The outburst stepped into decay phase after the transition from HSS to HIMS.

(\textit{b}) In the second outburst, two similar breaks took place around MJD 55424 and MJD 55454. Before MJD 55424, the fractional rms stayed on a high level above 20\%, and the characteristic frequency of QPOs increased with flux. Photon index increased to 2.1 before a sudden rise of thermal disk component emission. In the following, the rms fluctuated at a low level about a few percentages, and thermal disk component dominated the luminosity. Until MJD 55454, another similiar break appeared, which stepped into the decay phase: the PDS presented detectable timing noise with a rms amplitude of ~14\%, accompanied with an obviously increase of power-law component and a decreasing index. QPOs appeared two days later, with a decreasing frequency. The spectral and timing properties are expected for HIMS in decay phase. After the transition, the disk component became weaker and then non-observable.

(\textit{c}) We apply the same approach to the third outburst, and locate the transition between MJD 55672 and MJD 55687. Due to the absence of observations around MJD 55687, the break can not be precisely determinted, but in a range between MJD 55687 and MJD 55690.

\section{DISCUSSION AND SUMMARY}

In this paper we have studied three outbursts of BH XRB \H1743 observed by {\it RXTE} in its latest  service phase. Among them  outbursts 2010b and 2011 showed up in {\it RXTE} data with relatively complete evolution. The timing and spectral analysis  show that these two outbursts fulfilled the transition to a high-soft state. Therefore the number of outbursts recorded by {\it RXTE} with a relatively complete evolution profile increases to 5, which makes \H1743 among the few BH XRBs with sufficient outbursts be fully traced at X-rays.

\begin{figure}
\centering
	\includegraphics[width=0.5\textwidth]{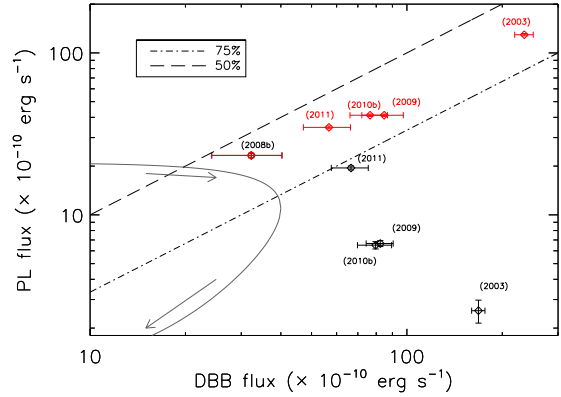}
		\caption{The disk component flux versus the power-law component flux. The semi-circle in solid line shows a clockwise evolution of the outburst in this diagram.  The dished line and the dish-dotted line denote the percentages of disk flux in total flux.
 (\textit{a}) Data   in black   are the leftmost   (the softest region) in HID of 4 outbursts 2003, 2009, 2010b and 2011. (\textit{b}) Data in red stands for where the first type-B QPO shows up in  outbursts except the 2008b. The data point for outburst 2008b is the lestmost in its HID.}
	\label{flux}
\end{figure}

Including a failed outburst 2008b, the enlarged complete outburst sample for \H1743 allows us to investigate the spectral transitions in more details. The comparison of 2008b to  two other outbursts 2003 and 2009 had already been done in Chen et al. (2010). It was reported that in the luminosity diagram of the corona versus the disk,  while the tracks of the outbursts in 2003 and 2009 cross the line which represents a roughly equal contribution to the emission from the thermal and the non-thermal components, the track of the 2008 outburst has its turn-over falling on this line. This may indicate an emission balance between the corona and the disk that prevents the state transition from going further beyond the low-hard state.  With two more outbursts 2010b and 2011, we investigated such a possibility further in Fig.~\ref{flux}, where the power-law flux is plotted against the disk black body component. In this diagram, the evolution of the outburst draws a clockwise semi-circle (see Fig.~\ref{flux}). We picked up for each outburst two most interesting data groups in evolution, i.e., Group One for data when the first type-B QPO shows up, representing usually the state transition to a high-soft state, and Group Two for data corresponding to the left most in HID. The flux ratio of  the thermal to the total emissions constitutes a region in this diagram if one takes the flux  ratios of  50\% and 75\% (Chen et al. 2010, Remillard et al. 2006). We find that  data in Group One are located in this region, while the data in Group Two well across it to step into a high-soft state, except 2008b (absence of type-B QPO). This is in support of Chen et al. (2010) and suggests a  balance between the thermal and non-thermal emission, which is most likely responsible for the state transition from a low-hard state to high-soft state, could be tightly constrained to within a range of  $\sim$ 50\%-75\%.

\begin{figure*}
\centering
\includegraphics[width=0.9\textwidth]{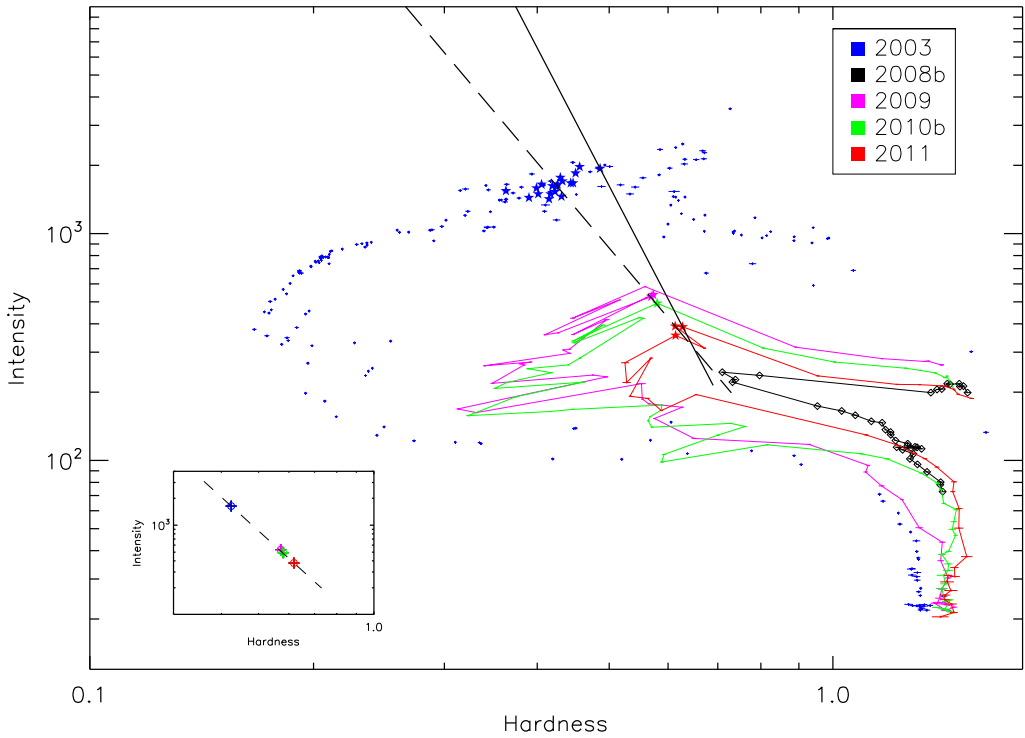}
\caption{The HIDs of the five complete outbursts. The hardness is defined with the ratio of count rates of the 6-15 keV and 3-6 keV. And the intensity is the total count rates in the 3-30 keV. The five point stars stand for the type-B QPOs. The solid line is the right boundary of the region where the type-B QPOs exist, while the dash line fit the mean value of the type-B QPOs of each outburst. The inset emphasize the four points fitting of mean value of type-B QPOs, in which we can see a perfect power-law distribution with a index of -3.3$\pm$0.2. Different colors accordingly belong to different outbursts.}
\label{5qtrack}
\end{figure*}

We investigated the 2008b failed outburst further in the context of the enlarged complete outburst sample of \H1743. We show the data points with type-B QPOs for each outburst in HID. As shown in Fig.~\ref{5qtrack}, these data are symbolized as stars of different colors, with which two lines are made accordingly. The solid line is the connection of the  stars that firstly showed up in 2003 outburst and 2011 outburst along  evolution of their q-tracks, which stands for the right boundary of the type-B QPOs region. The dashed line shows a linear fit in logarithm domain  for those star-symbolized data averaged over each outburst. The fit gives a power-law index of -3.3$\pm$0.2 under a reduced $\chi^2$ of 0.9 (1 dof) (see the inset in Fig.~\ref{5qtrack}). These two lines represent the spectral transitions to high-soft state for four outbursts and constitute a region in HID, an extension of which locates in the vicinity of the 2008 failed outburst, and exclude it from a successful group. The distance of the left most point in 2008b failed outburst to the extension of the dashed line is as small as roughly 0.02 on hardness scale.  We therefore suggest the 2008b, the failed outburst, may turn out to be the first ``almost successful outburst" (if the region is universal for a larger sample or all outbursts of this source, and the failed outburst would has crossed it, then it would stepped into high-soft state), which is in turn helpful for modelling the outburst, especially upon the condition required for transition to high-soft state.

In summary, with two more outbursts recorded almost completely by {\it RXTE} in its latest service phase we can constitute for the first time a rather sufficient outburst sample for \H1743. We put the 2008b failed outburst into this context and found that it might present, so far known to BH XRBs,  the first failed outburst experiencing almost successfully a  high-soft state.

\section*{ACKNOWLEDGEMENTS}

We thank the anonymous referee for insightful suggestions and Dr. Yizhong Fan for help in improving the presentation. This work is supported by the CAS key Project KJCX2-YW-T03, 973 program 2009CB824800, XTP project XDA04060604 and NSFC-11103020, 11133002, 11073021, 11173024, 11273064 and 11003045, This research has made use of data obtained from the High Energy Astrophysics Science Archive Research Center (HEASARC), provided by NASA's Goddard Space Flight Center.

\clearpage

\begin{figure*}
\centering
	\includegraphics[width=0.9\textwidth]{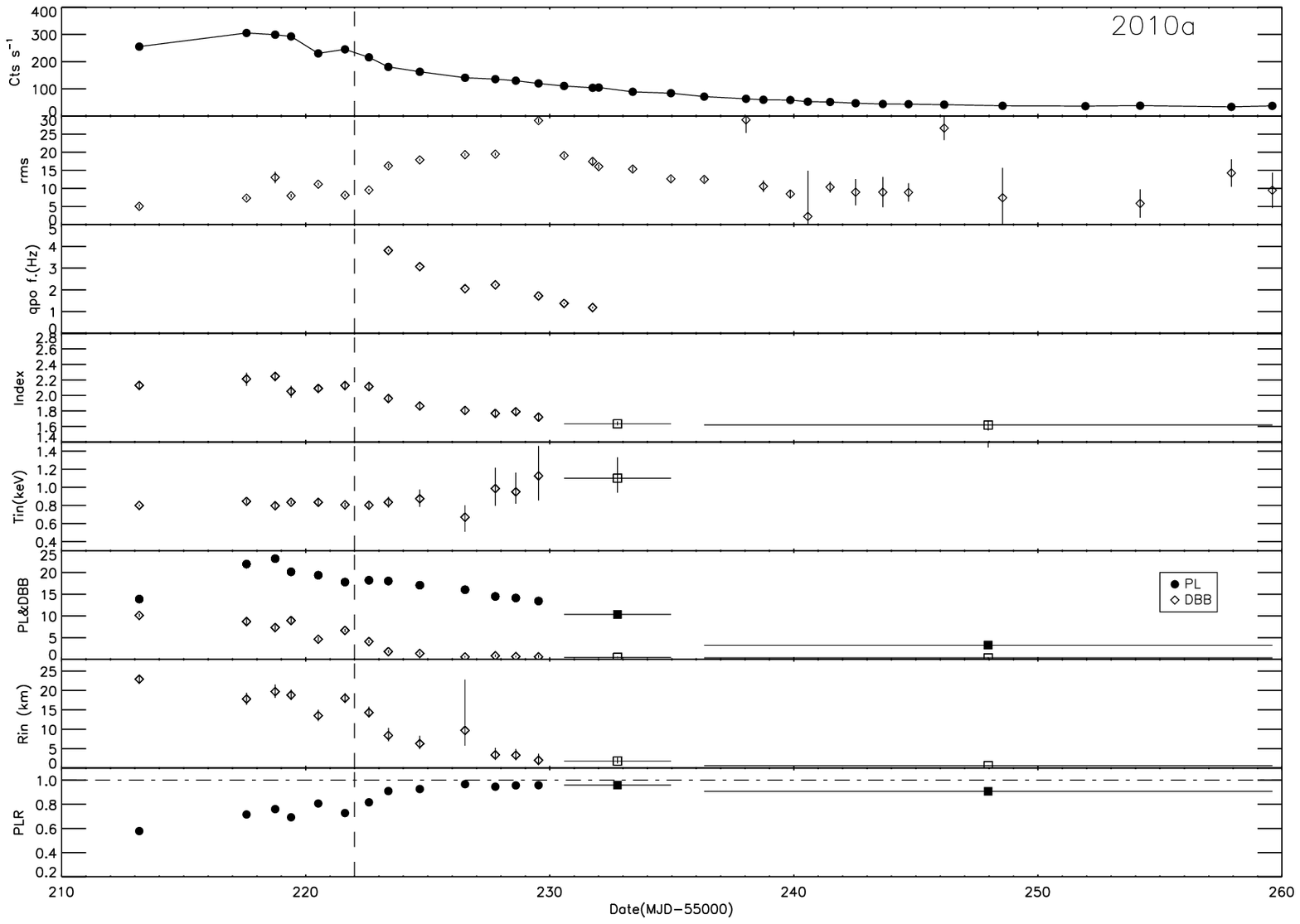}
	\includegraphics[width=0.9\textwidth]{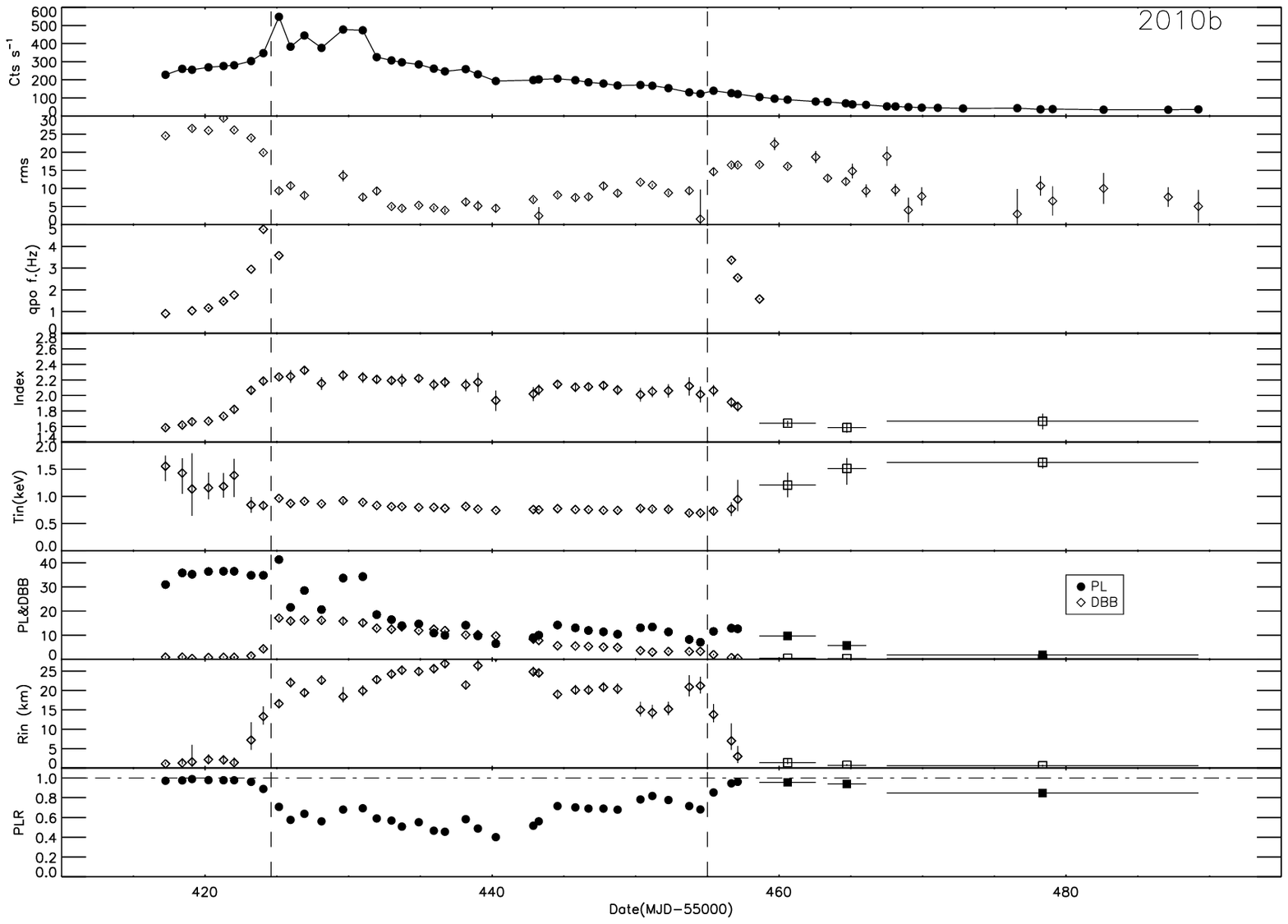}
\end{figure*}

\clearpage

\begin{table*}
\tiny
\caption{Spectral fitting results on outburst 2010a of \H1743}
\label{2010apara}
\begin{tabular}{cccccccccccc}
\hline \hline
ObsID &  Date & Rate & Tin & Rin & $\Gamma$ & DBB flux & PL flux & QPO Fre. & $\chi_{red}^{2}(d.o.f)$  \\
      &  &      &   &  &    &$\times$ 10$^{-10}$ &$\times$ 10$^{-10}$ &   & &\\
      &  (MJD)  &      & (keV) & (km)  &    & (ergs cm$^{-2}$ s$^{-1}$)& (ergs cm$^{-2}$ s$^{-1}$)& (Hz) & &\\
\hline
 95405-01-01-00 & 55213.179 & 255.3 & 0.80$^{+0.01}_{-0.01}$ &22.9$^{+1.1}_{-1.1 }$ & 2.13$^{+0.06}_{-0.06}$ & 10.12$^{+0.36}_{-0.34}$ & 13.87$^{+0.39}_{-0.44}$ &              	  &0.82(49)\\				
 95405-01-01-01 & 55217.574 & 305.5 & 0.85$^{+0.03}_{-0.03}$ &17.8$^{+1.5}_{-1.6 }$ & 2.21$^{+0.08}_{-0.09}$ &  8.71$^{+0.88}_{-0.84}$ & 21.93$^{+0.98}_{-1.08}$ &              	  &1.22(49)\\				
 95405-01-02-04 & 55218.748 & 299.2 & 0.80$^{+0.03}_{-0.03}$ &19.7$^{+1.6}_{-1.8 }$ & 2.25$^{+0.05}_{-0.06}$ &  7.33$^{+0.61}_{-0.55}$ & 23.17$^{+0.59}_{-0.74}$ &              	  &0.89(49)\\			
 95405-01-02-00 & 55219.399 & 292.8 & 0.84$^{+0.02}_{-0.02}$ &18.8$^{+1.3}_{-1.4 }$ & 2.05$^{+0.07}_{-0.08}$ &  8.95$^{+0.64}_{-0.60}$ & 20.14$^{+0.76}_{-0.77}$ &              	  &0.91(49)\\			
 95405-01-02-01 & 55220.513 & 230.3 & 0.84$^{+0.04}_{-0.03}$ &13.5$^{+1.4}_{-1.5 }$ & 2.09$^{+0.05}_{-0.06}$ &  4.65$^{+0.44}_{-0.43}$ & 19.36$^{+0.45}_{-0.53}$ &              	  &0.73(49)\\			
 95405-01-02-05 & 55221.609 & 245.2 & 0.81$^{+0.02}_{-0.02}$ &18.0$^{+1.1}_{-1.3 }$ & 2.13$^{+0.05}_{-0.05}$ &  6.67$^{+0.38}_{-0.37}$ & 17.78$^{+0.46}_{-0.44}$ &              	  &0.98(49)\\			
 95405-01-02-03 & 55222.590 & 215.9 & 0.80$^{+0.03}_{-0.03}$ &14.3$^{+1.2}_{-1.5 }$ & 2.12$^{+0.04}_{-0.05}$ &  4.10$^{+0.34}_{-0.33}$ & 18.20$^{+0.38}_{-0.46}$ &              	  &0.77(49)\\			
 95405-01-02-06 & 55223.385 & 180.6 & 0.84$^{+0.06}_{-0.06}$ &8.4 $^{+1.5}_{-1.9 }$ & 1.96$^{+0.05}_{-0.05}$ &  1.79$^{+0.32}_{-0.31}$ & 18.03$^{+0.33}_{-0.41}$ & 3.81$^{+0.04}_{-0.04}$ &0.84(49)\\
 95405-01-02-02 & 55224.673 & 162.7 & 0.87$^{+0.10}_{-0.09}$ &6.3 $^{+1.4}_{-2.0 }$ & 1.87$^{+0.05}_{-0.06}$ &  1.36$^{+0.36}_{-0.31}$ & 17.06$^{+0.36}_{-0.43}$ & 3.07$^{+0.03}_{-0.03}$ &1.02(49)\\
 95405-01-03-00 & 55226.528 & 140.7 & 0.67$^{+0.13}_{-0.16}$ &9.7 $^{+3.9}_{-13.1}$ & 1.81$^{+0.05}_{-0.05}$ &  0.56$^{+0.23}_{-0.22}$ & 16.03$^{+0.26}_{-0.29}$ & 2.06$^{+0.03}_{-0.03}$ &0.86(49)\\
 95405-01-03-04 & 55227.769 & 135.6 & 0.99$^{+0.23}_{-0.19}$ &3.4 $^{+1.3}_{-1.8 }$ & 1.77$^{+0.05}_{-0.05}$ &  0.83$^{+0.26}_{-0.26}$ & 14.49$^{+0.24}_{-0.26}$ & 2.23$^{+0.03}_{-0.03}$ &0.83(48)\\
 95405-01-03-01 & 55228.608 & 129.9 & 0.95$^{+0.21}_{-0.13}$ &3.3 $^{+1.3}_{-1.6 }$ & 1.79$^{+0.04}_{-0.04}$ &  0.65$^{+0.19}_{-0.18}$ & 14.13$^{+0.20}_{-0.19}$ &              	  &0.81(49)\\
 95405-01-03-05 & 55229.539 & 119.8 & 1.13$^{+0.33}_{-0.27}$ &2.0 $^{+0.8}_{-1.7 }$ & 1.72$^{+0.04}_{-0.06}$ &  0.59$^{+0.41}_{-0.24}$ & 13.43$^{+0.22}_{-0.34}$ & 1.72$^{+0.04}_{-0.03}$ &0.76(49)\\ \hline
 following 20   & 55232.775 & 98.3  & 1.10$^{+0.23}_{-0.16}$ &1.8 $^{+0.9}_{-0.6 }$ & 1.63$^{+0.03}_{-0.02}$ &  0.44$^{+0.15}_{-0.09}$ & 10.33$^{+0.08}_{-0.08}$ &              	  &0.77(49)\\
 OBSIDs (2 bins)& 55247.966 & 47.7  & 1.58$^{+0.11}_{-0.14}$ &0.6 $^{+0.1}_{-0.1 }$ & 1.62$^{+0.06}_{-0.07}$ &  0.34$^{+0.11}_{-0.10}$ &  3.27$^{+0.08}_{-0.09}$ &              	  &1.61(43)\\

\hline
\end{tabular}
\end{table*}

\clearpage

\begin{table*}
\tiny
\caption{Spectral fitting results on outburst 2010b of \H1743}
\label{2010bpara}
\begin{tabular}{cccccccccccc}
\hline \hline
ObsID &  MJD & Rate & Tin & Rin & $\Gamma$ & DBB flux & PL flux & QPO Fre. & $\chi_{red}^{2}(d.o.f)$  \\
      &  &      &   &  &    &$\times$ 10$^{-10}$ &$\times$ 10$^{-10}$ &   & &\\
      &  (MJD)  &      & (keV) & (km)  &    & (ergs cm$^{-2}$ s$^{-1}$)& (ergs cm$^{-2}$ s$^{-1}$)& (Hz) & &\\
\hline

 95368-01-01-00  & 55417.238 & 227.51 & 1.56$^{+0.19}_{-0.27}$ &1.1 $^{+0.2}_{-0.4}$ & 1.59$^{+0.02}_{-0.02}$ &  0.93$^{+0.42}_{-0.34}$ & 30.97$^{+0.29}_{-0.36}$& 0.91$^{+0.01}_{-0.01}$ &0.72(49) \\
 95360-14-01-00  & 55418.405 & 260.14 & 1.43$^{+0.27}_{-0.38}$ &1.3 $^{+0.3}_{-1.2}$ & 1.62$^{+0.02}_{-0.03}$ &  0.92$^{+0.57}_{-0.39}$ & 35.81$^{+0.33}_{-0.49}$&      		  &0.74(49) \\
 95360-14-02-01  & 55419.088 & 255.05 & 1.14$^{+0.65}_{-0.49}$ &1.6 $^{+1.2}_{-4.4}$ & 1.66$^{+0.03}_{-0.03}$ &  0.41			 & 35.24$^{+0.41}_{-0.48}$& 1.04$^{+0.01}_{-0.01}$ &1.25(49) \\
 95360-14-02-00  & 55420.233 & 268.71 & 1.16$^{+0.28}_{-0.21}$ &2.2 $^{+0.8}_{-1.3}$ & 1.67$^{+0.02}_{-0.02}$ &  0.82$^{+0.35}_{-0.24}$ & 36.39$^{+0.25}_{-0.25}$& 1.17$^{+0.00}_{-0.00}$ &1.04(49) \\
 95360-14-03-00  & 55421.281 & 275.69 & 1.18$^{+0.25}_{-0.21}$ &2.1 $^{+0.7}_{-1.2}$ & 1.73$^{+0.02}_{-0.02}$ &  0.86$^{+0.36}_{-0.26}$ & 36.48$^{+0.25}_{-0.25}$& 1.48$^{+0.00}_{-0.00}$ &1.19(49) \\
 95360-14-02-03  & 55422.023 & 280.15 & 1.39$^{+0.30}_{-0.40}$ &1.4 $^{+0.5}_{-1.3}$ & 1.82$^{+0.03}_{-0.04}$ &  0.88			 & 36.48$^{+0.51}_{-0.67}$& 1.77$^{+0.01}_{-0.01}$ &1.03(49) \\
 95360-14-02-02  & 55423.204 & 303.15 & 0.85$^{+0.14}_{-0.15}$ &7.2 $^{+2.5}_{-4.6}$ & 2.07$^{+0.03}_{-0.05}$ &  1.45$^{+0.62}_{-0.50}$ & 34.83$^{+0.49}_{-0.71}$& 2.95$^{+0.02}_{-0.01}$ &1.15(48) \\
 95360-14-03-01  & 55424.056 & 346.73 & 0.83$^{+0.06}_{-0.06}$ &13.3$^{+2.1}_{-2.6}$ & 2.19$^{+0.05}_{-0.05}$ &  4.36$^{+0.79}_{-0.73}$ & 34.83$^{+0.89}_{-0.87}$& 4.79$^{+0.03}_{-0.03}$ &0.80(49) \\
 95360-14-04-00  & 55425.144 & 547.55 & 0.97$^{+0.02}_{-0.02}$ &16.6$^{+0.9}_{-0.9}$ & 2.24$^{+0.03}_{-0.03}$ & 17.12$^{+0.66}_{-0.61}$ & 41.30$^{+0.58}_{-0.75}$& 3.58$^{+0.03}_{-0.03}$ &1.16(49) \\
 95360-14-02-04  & 55425.952 & 382.44 & 0.87$^{+0.02}_{-0.02}$ &22.0$^{+1.2}_{-1.3}$ & 2.25$^{+0.08}_{-0.08}$ & 15.89$^{+0.86}_{-0.83}$ & 21.53$^{+0.96}_{-1.02}$&      		  &1.43(49) \\
 95360-14-02-05  & 55426.918 & 444.16 & 0.91$^{+0.02}_{-0.02}$ &19.4$^{+0.9}_{-1.0}$ & 2.32$^{+0.04}_{-0.04}$ & 16.28$^{+0.67}_{-0.62}$ & 28.51$^{+0.66}_{-0.78}$&      		  &0.82(49) \\
 95360-14-05-00  & 55428.110 & 375.99 & 0.86$^{+0.02}_{-0.02}$ &22.6$^{+1.2}_{-1.3}$ & 2.16$^{+0.08}_{-0.09}$ & 16.15$^{+0.80}_{-0.77}$ & 20.61$^{+0.92}_{-0.93}$&      		  &1.15(49) \\
 95360-14-06-00  & 55429.615 & 477.29 & 0.92$^{+0.03}_{-0.04}$ &18.4$^{+1.5}_{-2.5}$ & 2.26$^{+0.06}_{-0.07}$ & 15.87$^{+1.12}_{-1.12}$ & 33.65$^{+1.18}_{-1.37}$&      		  &1.40(48) \\
 95360-14-07-00  & 55430.986 & 473.03 & 0.89$^{+0.02}_{-0.03}$ &19.9$^{+1.3}_{-1.4}$ & 2.23$^{+0.07}_{-0.07}$ & 15.18$^{+1.19}_{-1.14}$ & 34.28$^{+1.37}_{-1.47}$&      		  &0.59(49) \\
 95360-14-05-01  & 55431.959 & 324.93 & 0.83$^{+0.01}_{-0.02}$ &22.8$^{+1.1}_{-1.1}$ & 2.21$^{+0.06}_{-0.06}$ & 12.89$^{+0.50}_{-0.49}$ & 18.54$^{+0.56}_{-0.55}$&      		  &1.01(49) \\
 95360-14-06-01  & 55432.985 & 306.79 & 0.81$^{+0.01}_{-0.01}$ &24.2$^{+1.1}_{-1.2}$ & 2.19$^{+0.06}_{-0.06}$ & 12.49$^{+0.46}_{-0.44}$ & 16.44$^{+0.46}_{-0.52}$&      		  &0.53(49) \\
 95360-14-07-01  & 55433.711 & 296.24 & 0.81$^{+0.01}_{-0.01}$ &25.2$^{+1.1}_{-1.2}$ & 2.20$^{+0.08}_{-0.09}$ & 13.52$^{+0.54}_{-0.51}$ & 13.93$^{+0.59}_{-0.66}$&      		  &0.77(49) \\
 95360-14-07-02  & 55434.887 & 284.64 & 0.80$^{+0.01}_{-0.01}$ &24.9$^{+1.1}_{-1.1}$ & 2.22$^{+0.06}_{-0.06}$ & 11.93$^{+0.40}_{-0.37}$ & 14.69$^{+0.41}_{-0.43}$&      		  &0.96(49) \\
 95360-14-08-00  & 55435.937 & 261.23 & 0.80$^{+0.01}_{-0.01}$ &25.6$^{+1.0}_{-1.0}$ & 2.14$^{+0.07}_{-0.07}$ & 12.52$^{+0.36}_{-0.34}$ & 10.91$^{+0.38}_{-0.40}$&      		  &0.78(49) \\
 95360-14-09-00  & 55436.710 & 246.22 & 0.78$^{+0.01}_{-0.01}$ &26.9$^{+1.0}_{-1.0}$ & 2.17$^{+0.07}_{-0.07}$ & 11.93$^{+0.31}_{-0.29}$ &  9.98$^{+0.33}_{-0.32}$&      		  &0.65(49) \\
 95360-14-10-00  & 55438.161 & 258.58 & 0.82$^{+0.02}_{-0.02}$ &21.4$^{+1.1}_{-1.2}$ & 2.14$^{+0.08}_{-0.08}$ & 10.16$^{+0.48}_{-0.48}$ & 14.16$^{+0.57}_{-0.57}$&      		  &1.01(49) \\
 95360-14-11-00  & 55439.009 & 229.81 & 0.77$^{+0.02}_{-0.02}$ &26.4$^{+1.4}_{-1.5}$ & 2.17$^{+0.12}_{-0.13}$ & 10.23$^{+0.51}_{-0.52}$ &  9.75$^{+0.60}_{-0.63}$&      		  &0.86(49) \\
 95360-14-12-00  & 55440.258 & 192.73 & 0.74$^{+0.01}_{-0.01}$ &28.7$^{+1.4}_{-1.5}$ & 1.94$^{+0.13}_{-0.14}$ &  9.73$^{+0.29}_{-0.30}$ &  6.50$^{+0.34}_{-0.34}$&      		  &0.98(49) \\
 95360-14-13-00  & 55442.870 & 198.06 & 0.76$^{+0.01}_{-0.01}$ &24.8$^{+1.2}_{-1.3}$ & 2.02$^{+0.09}_{-0.09}$ &  8.41$^{+0.30}_{-0.31}$ &  8.97$^{+0.38}_{-0.40}$&      		  &1.06(49) \\
 95360-14-14-00  & 55443.250 & 201.95 & 0.75$^{+0.01}_{-0.01}$ &24.5$^{+1.1}_{-1.2}$ & 2.07$^{+0.07}_{-0.07}$ &  7.87$^{+0.28}_{-0.28}$ & 10.05$^{+0.33}_{-0.32}$&      		  &0.72(49) \\
 95360-14-15-00  & 55444.558 & 205.64 & 0.78$^{+0.02}_{-0.02}$ &19.0$^{+1.1}_{-1.2}$ & 2.14$^{+0.05}_{-0.05}$ &  5.67$^{+0.29}_{-0.28}$ & 14.22$^{+0.33}_{-0.36}$&      		  &0.75(49) \\
 95360-14-16-00  & 55445.800 & 197.46 & 0.76$^{+0.02}_{-0.02}$ &20.1$^{+1.2}_{-1.3}$ & 2.11$^{+0.05}_{-0.05}$ &  5.54$^{+0.28}_{-0.27}$ & 13.03$^{+0.30}_{-0.33}$&      		  &1.18(49) \\
 95360-14-17-00  & 55446.715 & 185.85 & 0.76$^{+0.02}_{-0.02}$ &20.1$^{+1.2}_{-1.3}$ & 2.11$^{+0.05}_{-0.05}$ &  5.38$^{+0.26}_{-0.25}$ & 11.97$^{+0.28}_{-0.30}$&      		  &1.32(49) \\
 95360-14-18-00  & 55447.758 & 179.13 & 0.74$^{+0.02}_{-0.02}$ &20.8$^{+1.2}_{-1.3}$ & 2.13$^{+0.05}_{-0.06}$ &  5.13$^{+0.25}_{-0.24}$ & 11.40$^{+0.27}_{-0.31}$&      		  &1.06(49) \\
 95360-14-19-00  & 55448.743 & 168.38 & 0.74$^{+0.02}_{-0.02}$ &20.4$^{+1.3}_{-1.4}$ & 2.07$^{+0.06}_{-0.07}$ &  4.93$^{+0.25}_{-0.25}$ & 10.42$^{+0.29}_{-0.33}$&      		  &0.99(49) \\
 95360-14-20-00  & 55450.334 & 170.98 & 0.78$^{+0.04}_{-0.04}$ &15.0$^{+1.7}_{-2.1}$ & 2.01$^{+0.09}_{-0.09}$ &  3.64$^{+0.43}_{-0.43}$ & 13.06$^{+0.52}_{-0.53}$&      		  &0.68(49) \\
 95360-14-20-01  & 55451.165 & 166.97 & 0.77$^{+0.04}_{-0.04}$ &14.3$^{+1.6}_{-2.0}$ & 2.05$^{+0.07}_{-0.07}$ &  3.00$^{+0.37}_{-0.35}$ & 13.43$^{+0.41}_{-0.46}$&      		  &0.86(49) \\
 95360-14-21-00  & 55452.286 & 153.84 & 0.76$^{+0.04}_{-0.04}$ &15.2$^{+1.6}_{-1.9}$ & 2.06$^{+0.08}_{-0.09}$ &  3.28$^{+0.39}_{-0.37}$ & 11.38$^{+0.43}_{-0.49}$&      		  &1.27(49) \\
 95360-14-21-01  & 55453.736 & 130.50 & 0.70$^{+0.03}_{-0.03}$ &20.9$^{+2.4}_{-3.0}$ & 2.12$^{+0.11}_{-0.12}$ &  3.28$^{+0.37}_{-0.35}$ &  8.24$^{+0.37}_{-0.44}$&      		  &0.79(49) \\
 95360-14-22-00  & 55454.511 & 122.66 & 0.69$^{+0.02}_{-0.03}$ &21.2$^{+2.0}_{-2.3}$ & 2.02$^{+0.10}_{-0.10}$ &  3.32$^{+0.24}_{-0.25}$ &  7.10$^{+0.28}_{-0.29}$&      		  &0.64(49) \\
 95360-14-22-01  & 55455.429 & 139.39 & 0.73$^{+0.05}_{-0.05}$ &13.8$^{+2.0}_{-2.7}$ & 2.06$^{+0.07}_{-0.08}$ &  2.00$^{+0.31}_{-0.30}$ & 11.59$^{+0.35}_{-0.37}$& 3.37$^{+0.06}_{-0.05}$ &0.79(49) \\
 95360-14-23-00  & 55456.673 & 125.95 & 0.77$^{+0.12}_{-0.13}$ &7.0 $^{+2.3}_{-4.5}$ & 1.91$^{+0.06}_{-0.07}$ &  0.75$^{+0.28}_{-0.26}$ & 12.91$^{+0.27}_{-0.35}$& 2.56$^{+0.05}_{-0.04}$ &0.62(49) \\
 95360-14-23-01  & 55457.115 & 120.80 & 0.94$^{+0.36}_{-0.21}$ &3.0 $^{+1.7}_{-2.7}$ & 1.86$^{+0.06}_{-0.07}$ &  0.51$^{+0.30}_{-0.25}$ & 12.65$^{+0.24}_{-0.29}$& 1.58$^{+0.04}_{-0.03}$ &0.67(49) \\ \hline
 following 20    & 55460.591 & 92.40  & 1.21$^{+0.23}_{-0.22}$ &1.4 $^{+0.9}_{-0.4}$ & 1.64$^{+0.03}_{-0.04}$ &  0.45$^{+0.20}_{-0.13}$ &  9.67$^{+0.11}_{-0.09}$&      		  &1.07(49) \\
 OBSIDs (3 bins) & 55464.720 & 68.21  & 1.51$^{+0.19}_{-0.30}$ &0.7 $^{+0.3}_{-0.1}$ & 1.59$^{+0.06}_{-0.07}$ &  0.37$^{+0.18}_{-0.16}$ &  5.73$^{+0.13}_{-0.16}$&      		  &0.99(48) \\
                 & 55478.366 & 42.57  & 1.63$^{+0.09}_{-0.11}$ &0.6 $^{+0.0}_{-0.1}$ & 1.67$^{+0.10}_{-0.11}$ &  0.33$^{+0.09}_{-0.09}$ &  1.80$^{+0.08}_{-0.08}$&      		  &1.36(48) \\
\hline
\end{tabular}
\end{table*}

\clearpage

\begin{table*}
\tiny
\caption{Spectral fitting results on outburst 2011 of \H1743}
\label{2011para}
\begin{tabular}{cccccccccccc}
\hline \hline
ObsID &  MJD & Rate & Tin & Rin & $\Gamma$ & DBB flux & PL flux & QPO Fre. & $\chi_{red}^{2}(d.o.f)$  \\
      &  &      &   &  &    &$\times$ 10$^{-10}$ &$\times$ 10$^{-10}$ &   & &\\
      &  (MJD)  &      & (keV) & (km)  &    & (ergs cm$^{-2}$ s$^{-1}$)& (ergs cm$^{-2}$ s$^{-1}$)& (Hz) & &\\
\hline

 96425-01-01-00 & 55663.669 & 209.2 & 1.46$^{+0.23}_{-0.28}$ &1.3 $^{+0.3}_{-0.6}$ & 1.49$^{+0.03}_{-0.03}$ &  3.92$^{+1.65}_{-1.25}$ & 29.21$^{+0.31}_{-0.42}$ & 0.43$^{+0.00}_{-0.01}$ & 0.63(49)\\		
 96425-01-01-01 & 55665.894 & 217.7 & 1.55$^{+0.17}_{-0.29}$ &1.4 $^{+0.2}_{-0.7}$ & 1.49$^{+0.03}_{-0.04}$ &  1.63$^{+0.58}_{-0.54}$ & 29.66$^{+0.45}_{-0.53}$ & 0.53$^{+0.00}_{-0.00}$ & 0.87(49)\\
 96425-01-02-00 & 55667.587 & 228.7 & 1.32$^{+0.23}_{-0.23}$ &1.5 $^{+0.4}_{-0.7}$ & 1.59$^{+0.02}_{-0.02}$ &  0.77$^{+0.32}_{-0.22}$ & 31.45$^{+0.19}_{-0.27}$ & 0.67$^{+0.00}_{-0.00}$ & 0.85(49)\\
 96425-01-02-04 & 55668.466 & 234.5 & 1.30$^{+0.53}_{-0.34}$ &1.8 $^{+0.8}_{-1.7}$ & 1.60$^{+0.04}_{-0.05}$ &  1.05$^{+1.12}_{-0.48}$ & 31.72$^{+0.48}_{-0.84}$ &                        & 0.90(49)\\
 96425-01-02-01 & 55668.978 & 235.4 & 1.31$^{+0.29}_{-0.30}$ &1.7 $^{+0.5}_{-1.3}$ & 1.63$^{+0.03}_{-0.04}$ &  0.98$^{+0.60}_{-0.39}$ & 32.11$^{+0.35}_{-0.53}$ & 0.89$^{+0.01}_{-0.01}$ & 0.93(49)\\
 96425-01-02-02 & 55670.619 & 239.7 & 0.98$^{+0.20}_{-0.15}$ &3.2 $^{+1.2}_{-1.8}$ & 1.75$^{+0.02}_{-0.02}$ &  0.69$^{+0.21}_{-0.18}$ & 31.13$^{+0.21}_{-0.22}$ & 1.33$^{+0.01}_{-0.01}$ & 1.21(49)\\
 96425-01-02-05 & 55671.524 & 240.4 & 1.03$^{+0.29}_{-0.19}$ &3.2 $^{+1.4}_{-2.1}$ & 1.83$^{+0.03}_{-0.03}$ &  0.91$^{+0.18}_{-0.11}$ & 30.07$^{+0.29}_{-0.32}$ & 1.83$^{+0.01}_{-0.01}$ & 0.77(49)\\
 96425-01-02-03 & 55672.841 & 262.6 & 0.77$^{+0.10}_{-0.11}$ &11.0$^{+3.2}_{-5.3}$ & 2.09$^{+0.05}_{-0.05}$ &  1.75$^{+0.56}_{-0.51}$ & 29.19$^{+0.57}_{-0.67}$ & 3.61$^{+0.03}_{-0.03}$ & 0.67(49)\\
 96425-01-03-00 & 55674.006 & 433.4 & 0.93$^{+0.02}_{-0.02}$ &15.6$^{+1.0}_{-1.1}$ & 2.21$^{+0.04}_{-0.04}$ & 11.67$^{+0.64}_{-0.61}$ & 34.69$^{+0.66}_{-0.73}$ & 3.53$^{+0.03}_{-0.03}$ & 0.78(49)\\
 96425-01-03-05 & 55674.275 & 434.6 & 0.93$^{+0.03}_{-0.03}$ &16.0$^{+1.2}_{-1.2}$ & 2.21$^{+0.05}_{-0.05}$ & 12.35$^{+0.84}_{-0.79}$ & 34.26$^{+0.89}_{-0.98}$ & 3.48$^{+0.03}_{-0.03}$ & 0.71(49)\\
 96425-01-03-01 & 55675.120 & 352.3 & 0.85$^{+0.03}_{-0.03}$ &16.9$^{+1.4}_{-1.4}$ & 2.19$^{+0.04}_{-0.04}$ &  7.85$^{+0.56}_{-0.54}$ & 29.79$^{+0.61}_{-0.65}$ &                        & 1.02(49)\\
 96425-01-03-02 & 55676.404 & 399.0 & 0.88$^{+0.02}_{-0.02}$ &17.7$^{+1.2}_{-1.2}$ & 2.19$^{+0.04}_{-0.05}$ & 11.13$^{+0.65}_{-0.62}$ & 31.16$^{+0.70}_{-0.75}$ & 3.30$^{+0.03}_{-0.03}$ & 1.12(49)\\
 96425-01-03-03 & 55678.055 & 309.6 & 0.83$^{+0.02}_{-0.02}$ &21.0$^{+1.1}_{-1.2}$ & 2.16$^{+0.05}_{-0.06}$ & 10.79$^{+0.47}_{-0.47}$ & 19.49$^{+0.54}_{-0.55}$ &                        & 0.72(49)\\
 96425-01-03-04 & 55679.293 & 255.0 & 0.80$^{+0.02}_{-0.02}$ &20.9$^{+1.2}_{-1.3}$ & 2.19$^{+0.06}_{-0.06}$ &  8.30$^{+0.43}_{-0.42}$ & 16.57$^{+0.49}_{-0.52}$ &                        & 1.10(49)\\
 96425-01-04-02 & 55680.224 & 319.8 & 0.86$^{+0.03}_{-0.03}$ &18.1$^{+1.4}_{-1.5}$ & 2.16$^{+0.08}_{-0.08}$ & 10.24$^{+0.79}_{-0.77}$ & 22.56$^{+0.89}_{-0.95}$ &                        & 1.24(49)\\
 96425-01-04-00 & 55681.172 & 223.3 & 0.78$^{+0.02}_{-0.02}$ &20.8$^{+1.2}_{-1.3}$ & 2.18$^{+0.05}_{-0.06}$ &  7.05$^{+0.35}_{-0.34}$ & 14.57$^{+0.40}_{-0.43}$ &                        & 0.83(49)\\
 96425-01-04-03 & 55682.183 & 217.7 & 0.76$^{+0.03}_{-0.03}$ &21.4$^{+2.1}_{-2.4}$ & 2.20$^{+0.10}_{-0.11}$ &  6.36$^{+0.67}_{-0.66}$ & 14.77$^{+0.76}_{-0.84}$ &                        & 0.98(49)\\
 96425-01-04-01 & 55684.657 & 194.2 & 0.77$^{+0.03}_{-0.02}$ &19.0$^{+1.4}_{-1.7}$ & 2.17$^{+0.07}_{-0.08}$ &  5.29$^{+0.42}_{-0.40}$ & 13.43$^{+0.47}_{-0.51}$ &                        & 0.87(49)\\
 96425-01-05-00 & 55687.591 & 224.6 & 0.78$^{+0.03}_{-0.03}$ &18.2$^{+1.5}_{-1.6}$ & 2.14$^{+0.05}_{-0.06}$ &  5.28$^{+0.41}_{-0.40}$ & 17.63$^{+0.47}_{-0.50}$ &                        & 0.81(49)\\
 96425-01-05-01 & 55690.131 & 149.5 & 0.88$^{+0.21}_{-0.17}$ &4.6 $^{+2.0}_{-3.4}$ & 1.93$^{+0.06}_{-0.06}$ &  0.75$^{+0.35}_{-0.31}$ & 16.19$^{+0.32}_{-0.38}$ & 3.03$^{+0.06}_{-0.06}$ & 1.36(49)\\
 96425-01-05-02 & 55691.512 & 134.5 & 0.96$^{+0.23}_{-0.16}$ &3.7 $^{+1.5}_{-2.2}$ & 1.76$^{+0.05}_{-0.06}$ &  0.80$^{+0.29}_{-0.26}$ & 15.12$^{+0.26}_{-0.33}$ & 2.02$^{+0.03}_{-0.03}$ & 0.77(49)\\
 96425-01-05-03 & 55693.000 & 125.2 & 1.29$^{+0.43}_{-0.52}$ &1.2 $^{+0.6}_{-2.5}$ & 1.77$^{+0.06}_{-0.09}$ &  0.49$^{+0.67}_{-0.36}$ & 14.21$^{+0.31}_{-0.59}$ & 1.82$^{+0.04}_{-0.03}$ & 0.71(49)\\
 96425-01-06-00 & 55694.003 & 119.8 & 0.84$^{+0.19}_{-0.17}$ &4.3 $^{+1.8}_{-3.6}$ & 1.68$^{+0.05}_{-0.06}$ &  0.51$^{+0.23}_{-0.20}$ & 13.42$^{+0.24}_{-0.29}$ & 1.51$^{+0.04}_{-0.02}$ & 1.04(49)\\
 96425-01-06-01 & 55695.422 & 109.9 & 0.87$^{+0.40}_{-0.23}$ &3.3 $^{+2.1}_{-4.3}$ & 1.68$^{+0.05}_{-0.06}$ &  0.35$^{+0.20}_{-0.18}$ & 12.55$^{+0.20}_{-0.23}$ & 1.20$^{+0.02}_{-0.02}$ & 1.12(49)\\ \hline
 following 15   & 55699.444 &  82.1 & 1.11$^{+0.78}_{-0.22}$ &1.5 $^{+1.1}_{-0.9}$ & 1.58$^{+0.04}_{-0.05}$ &  0.33$^{+0.51}_{-0.01}$ & 8.64 $^{+0.12}_{-0.16}$ &                        & 0.81(49)\\
 OBSIDs (2 bins)& 55720.048 &  41.1 & 1.77$^{+0.10}_{-0.11}$ &0.5 $^{+0.1}_{-0.1}$ & 1.58$^{+0.15}_{-0.17}$ &  0.41$^{+0.13}_{-0.12}$ & 1.68 $^{+0.11}_{-0.11}$ &                        & 1.38(48)\\

\hline
\end{tabular}
\end{table*}

\label{lastpage}

\end{document}